\documentclass{tlp}
\usepackage{mathptmx}
\usepackage{url}
\usepackage{color}
\usepackage{tabto}
\usepackage{amsmath}
\usepackage{pifont}
\usepackage{tikz}
\usepackage{pdfpages}
\usepackage[font=small,skip=0pt]{caption}

\usepackage{hyperref}

\title[Exception Handling is Backjumping]{Backjumping is Exception Handling}
  \author[E. Robbins, A. King and J. M. Howe]
	{ED ROBBINS, ANDY KING\\
	 University of Kent, Canterbury, CT2 7NF, UK
	 \and JACOB M. HOWE\\
	City, University of London, EC1V 0HB, UK}

\jdate{May 2017}
\pubyear{2017}
\submitted{15 September 2017}

\pagerange{\pageref{firstpage}--\pageref{lastpage}}
\doi{S1471068401001193}

\begin{document}

\label{firstpage}
\maketitle
  \begin{abstract}

ISO Prolog provides catch and throw to realise the control flow of
exception handling. This pearl demonstrates that catch and throw
are inconspicuously amenable to the implementation of backjumping.
In fact, they have precisely the semantics required: rewinding the
search to a specific point, and carrying of a preserved term to
that point.  The utility of these properties is demonstrated through
an implementation of graph colouring  with backjumping
and a backjumping SAT solver that applies Conflict Driven Clause Learning.

  \end{abstract}

\section{Introduction}

The ISO Prolog reference manual~\cite{prolog96standard} explains how catch and
throw can pass control from one point of the program to another. The default
behaviour of \verb|catch(Goal, Catcher, RecoveryGoal)| is to simply invoke
\verb|Goal|.  However, if during execution there is a call to \verb|throw(Ball)| then control
(and bindings) are unwound to the closest ancestor \verb|catch| in the call
stack which matches against the term \verb|Ball|.  Specifically, if the
\verb|Catcher| argument of the closest \verb|catch| unifies with a copy of \verb|Ball|,
then the \verb|RecoveryGoal| meta-call of that \verb|catch| is invoked.
Otherwise, control is unwound further until a matching \verb|catch| is found.
Since bindings are undone as the call stack is unwound, \verb|Ball| might also be
used to communicate information to \verb|RecoveryGoal|, for example to report
the nature of a failure.

The power of this control flow construct is that it can transfer control to a
specific point in the call stack by using the \verb|Ball| to target a specific
\verb|catch|. This is exactly what is required for backjumping. Backjumping
\cite{stallman77forward,gaschnig79performance}, in contrast to
chronological backtracking, leaps across multiple levels in a search tree
directly to the decision that triggered failure, rather than stepping through
each decision, one by one. Backjumping has found application in 
truth maintenance systems \cite{kleer86assumption}, 
logic programming \cite{bruynooghe80analysis}, 
constraint solving \cite{dechter90enhancement},
and most recently in SAT
to realise~\cite{marquessilva96grasp} Conflict Driven Clause Learning (CDCL). 
A CDCL solver requires not only search to be unwound to a specific decision (by
backjumping), but also a term (a learnt clause) to be preserved and carried to
that decision. Fortuitously this facility is also provided by catch and throw.

The problem of adding control to the logic of a search algorithm sits at the
very heart of logic programming~\cite{kowalski79algorithm}.
However, how control is added, and the clarity of the control component,
can be controversial. 
The problem of adding control is
particularly acute when programming
search problems like SAT, where the problem statement
can be very simple, but the best algorithms analyse the decisions to
focus search (as in CDCL). 
Moreover, backjumping is at
odds with chronological backtracking, and as a consequence certain classes of
algorithms that at first glance appear well-suited to logic programming
are, in fact, almost incompatible with the paradigm, at least in its purest form. 

This pearl proposes catch and
throw for programming backjumping, 
work that grew out of the
(irritating) problem of how to clearly code CDCL in Prolog. 
This chimes with
Bentley who coined the term programming pearl, and
wrote, ``Just as natural pearls grow from grains of sand
that have irritated oysters, these programming pearls have
grown from real problems that have irritated real programmers'' \cite{bentley86programming}.  
In contrast to a previous pearl \cite{bruynooghe04enchancing}, which used
a mutable database to orchestrate all aspects of intelligent backtracking, this paper
breaks down the problem of implementing CDCL into its various components which
are then matched against the language constructs of Prolog.  The net result is clarity.
To be precise, CDCL decomposes into 3 components:
(1) rewinding search and bindings, (2) communication of a newly learned clause
to its insertion point in the search tree, and (3) retaining learned clauses
(across backtracking and backjumping).   This paper argues that
catch and throw provide (1) and (2), whereas (3) is naturally provided by a
mutable database that might be implemented with a dynamic predicate, 
blackboard~\cite{bosschere93multi}, or non-backtrackable global
variables~\cite{wielemaker11swi}. 
Applications are not limited to SAT, or even SMT \cite{robbins15theory};
to demonstrate versatility the approach is first 
illustrated on the classic problem of graph colouring, providing a
template for backjumping with \verb|catch| and \verb|throw|, then, second it is applied
for SAT with learning.

Catch and throw have been advocated for programming backjumping
before as part of a \verb|comp.lang.prolog| discussion \cite{baljeu05backjump}, 
but the authors are not aware of any
studies which actually demonstrate the viability of the idea.  
Deploying catch and throw in backjumping
is unconventional since exception handling is intended to support
exceptional behaviour,
whereas in backjumping these constructs are used for the intended
control-flow, which is in turn exceptional in the context of Prolog's
execution model.  The discussion of \cite{baljeu05backjump} is centred on
the  use of these non-logical ISO language features in Prolog programming.

  

The rest of this paper is structured as two case studies on backjumping, the first on
graph colouring and the second on SAT.
The graph colouring study has been chosen as a minimal example of depth-first search
with backjumping, and the code provides a template for other examples. 
Section~\ref{sec:colour} 
explains how
\verb|catch| and \verb|throw| can be used to realise backjumping for a graph colouring problem, 
where the edge constraints are realised
as tests which check the colour assigned to each vertex as it becomes bound.
Section~\ref{sec:sat} moves onto SAT, building on the template provided by the graph colouring study
to illustrate how \verb|catch| and \verb|throw| 
can be deployed  to
communicate (learnt) information back to the \verb|catch|, necessary
when guiding search using CDCL.
Section~\ref{sec:discuss} presents the concluding discussion.




\section{Graph Colouring}\label{sec:colour}

The first of the two worked examples in this paper considers graph colouring, adding 
backjumping to depth-first search.  This example has been chosen as a minimal
illustrating example, with the code in Fig.~\ref{fig-colour-backjump} providing a template for other search problems, including
the motivating example of SAT solving with CDCL.

Figure~\ref{fig-animate} illustrates depth-first search with backjumping for a colouring problem,
where the objective is to assign red or green to each of the six vertices of
the graph so that vertices which share an edge are coloured differently.
The example uses just two colours for simplicity.
The vertices of the graph are ordered, as indicated by the numbering.
Colouring commences at vertex~1, red is tried before green at each vertex, and
white indicates the absence of a colour assignment. 
Each partially coloured
graph is augmented with a map which associates each vertex with a (possibly
empty) set of conflicts.  A conflict is a set of vertices with a 
colour assignment
that cannot be extended to satisfy all the edge constraints.

The conflict map is initially empty and is extended as each
vertex is coloured, and contracted on backtracking and backjumping. The first conflict occurs
in diagram 4 of Fig.~\ref{fig-animate} between vertices 1 and 3, and is recorded in the conflict map for
vertex~3; the conflict is always logged on the most recently assigned vertex.
For expositional purposes, 1, 3 are coloured red indicating
the partial colour assignment when the conflict is detected.

\def \radius {0.7cm}
\def \angle {360/6}
\def \diameter {4pt}
\def \starwidth {0.14 \textwidth} 
\def \mapwidth {0.17 \textwidth}

\begin{figure}[t!]
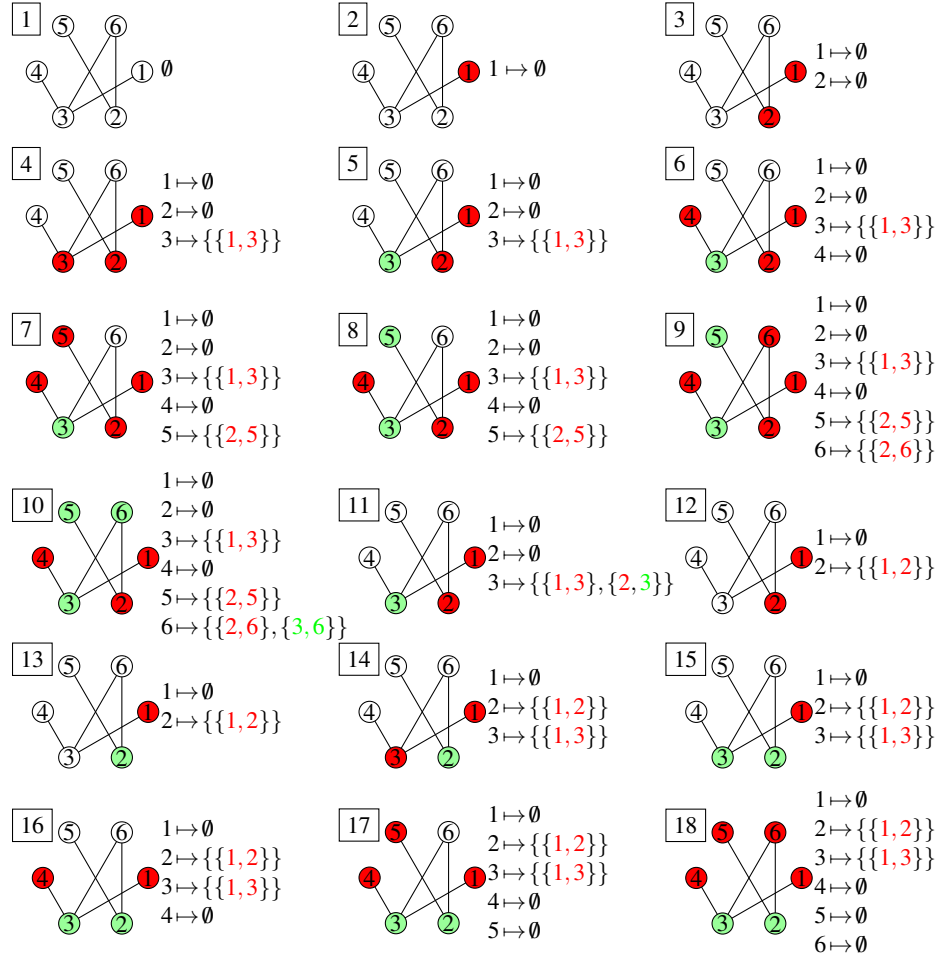

\begin{center}
\begin{tabular}{@{}c@{}c@{}c@{}c@{}c@{}c@{}}

\begin{minipage}{\starwidth}
\centering
\begin{tikzpicture}
\input{star-invariants}
\node[draw] at (-\radius*1.2,\radius) {1};
\filldraw [draw=black,fill=white] (v1) circle (\diameter);
\filldraw [draw=black,fill=white] (v2) circle (\diameter);
\filldraw [draw=black,fill=white] (v3) circle (\diameter);
\filldraw [draw=black,fill=white] (v4) circle (\diameter);
\filldraw [draw=black,fill=white] (v5) circle (\diameter);
\filldraw [draw=black,fill=white] (v6) circle (\diameter);

\foreach \vertex in {1,...,6}
{
  \node[thick] at ({360 - (\vertex - 1) * \angle}: {\radius * 1.0} ) {{\vertex}};
}
\end{tikzpicture}
\end{minipage}

&

\begin{minipage}{\mapwidth}
\begin{flushleft}
$\emptyset$
\end{flushleft}
\end{minipage}

&

\begin{minipage}{\starwidth}
\centering
\begin{tikzpicture}
\input{star-invariants}
\node[draw] at (-\radius*1.2,\radius) {2};
\filldraw [draw=black,fill=red] (v1) circle (\diameter);
\filldraw [draw=black,fill=white] (v2) circle (\diameter);
\filldraw [draw=black,fill=white] (v3) circle (\diameter);
\filldraw [draw=black,fill=white] (v4) circle (\diameter);
\filldraw [draw=black,fill=white] (v5) circle (\diameter);
\filldraw [draw=black,fill=white] (v6) circle (\diameter);

\foreach \vertex in {1,...,6}
{
  \node at ({360 - (\vertex - 1) * \angle}: {\radius * 1.0} ) {{\vertex}};
}

\end{tikzpicture}
\end{minipage}

&

\begin{minipage}{\mapwidth}
\begin{flushleft}
$
1 \mapsto \emptyset
$
\end{flushleft}
\end{minipage}

&

\begin{minipage}{\starwidth}
\centering
\begin{tikzpicture}
\input{star-invariants}
\node[draw] at (-\radius*1.2,\radius) {3};
\filldraw [draw=black,fill=red] (v1) circle (\diameter);
\filldraw [draw=black,fill=red] (v2) circle (\diameter);
\filldraw [draw=black,fill=white] (v3) circle (\diameter);
\filldraw [draw=black,fill=white] (v4) circle (\diameter);
\filldraw [draw=black,fill=white] (v5) circle (\diameter);
\filldraw [draw=black,fill=white] (v6) circle (\diameter);

\foreach \vertex in {1,...,6}
{
  \node at ({360 - (\vertex - 1) * \angle}: {\radius * 1.0} ) {{\vertex}};
}
\end{tikzpicture}
\end{minipage}

&

\begin{minipage}{\mapwidth}
\begin{flushleft}
$
\begin{array}{@{}r@{\,}c@{\,}l@{}}
1 & \mapsto  & \emptyset \\
2 & \mapsto  & \emptyset 
\end{array}
$
\end{flushleft}
\end{minipage}

\\[6ex]

\begin{minipage}{\starwidth}
\centering
\begin{tikzpicture}
\input{star-invariants}
\node[draw] at (-\radius*1.2,\radius) {4};
\filldraw [draw=black,fill=red] (v1) circle (\diameter);
\filldraw [draw=black,fill=red] (v2) circle (\diameter);
\filldraw [draw=black,fill=red] (v3) circle (\diameter);
\filldraw [draw=black,fill=white] (v4) circle (\diameter);
\filldraw [draw=black,fill=white] (v5) circle (\diameter);
\filldraw [draw=black,fill=white] (v6) circle (\diameter);

\foreach \vertex in {1,...,6}
{
  \node at ({360 - (\vertex - 1) * \angle}: {\radius * 1.0} ) {{\vertex}};
}
\end{tikzpicture}
\end{minipage}

&

\begin{minipage}{\mapwidth}
\begin{flushleft}
$
\begin{array}{@{}r@{\,}c@{\,}l@{}}
1 & \mapsto  & \emptyset \\
2 & \mapsto  & \emptyset \\
3 & \mapsto  & \{ \{ \textcolor{red}{1, 3} \} \} 
\end{array}
$
\end{flushleft}
\end{minipage}

&
\begin{minipage}{\starwidth}
\centering
\begin{tikzpicture}
\input{star-invariants}
\node[draw] at (-\radius*1.2,\radius) {5};
\filldraw [draw=black,fill=red] (v1) circle (\diameter);
\filldraw [draw=black,fill=red] (v2) circle (\diameter);
\filldraw [draw=black,fill=green!40] (v3) circle (\diameter);
\filldraw [draw=black,fill=white] (v4) circle (\diameter);
\filldraw [draw=black,fill=white] (v5) circle (\diameter);
\filldraw [draw=black,fill=white] (v6) circle (\diameter);

\foreach \vertex in {1,...,6}
{
  \node at ({360 - (\vertex - 1) * \angle}: {\radius * 1.0} ) {{\vertex}};
}
\end{tikzpicture}
\end{minipage}

&

\begin{minipage}{\mapwidth}
\begin{flushleft}
$
\begin{array}{@{}r@{\,}c@{\,}l@{}}
1 & \mapsto  & \emptyset \\
2 & \mapsto  & \emptyset \\
3 & \mapsto  & \{ \{ \textcolor{red}{1, 3} \} \} \\
\end{array}
$
\end{flushleft}
\end{minipage}

&

\begin{minipage}{\starwidth}
\centering
\begin{tikzpicture}
\input{star-invariants}
\node[draw] at (-\radius*1.2,\radius) {6};
\filldraw [draw=black,fill=red] (v1) circle (\diameter);
\filldraw [draw=black,fill=red] (v2) circle (\diameter);
\filldraw [draw=black,fill=green!40] (v3) circle (\diameter);
\filldraw [draw=black,fill=red] (v4) circle (\diameter);
\filldraw [draw=black,fill=white] (v5) circle (\diameter);
\filldraw [draw=black,fill=white] (v6) circle (\diameter);

\foreach \vertex in {1,...,6}
{
  \node at ({360 - (\vertex - 1) * \angle}: {\radius * 1.0} ) {{\vertex}};
}
\end{tikzpicture}
\end{minipage}

&

\begin{minipage}{\mapwidth}
\begin{flushleft}
$
\begin{array}{@{}r@{\,}c@{\,}l@{}}
1 & \mapsto  & \emptyset \\
2 & \mapsto  & \emptyset \\
3 & \mapsto  & \{ \{ \textcolor{red}{1, 3} \} \} \\
4 & \mapsto  & \emptyset
\end{array}
$
\end{flushleft}
\end{minipage}

\\[6ex]

\begin{minipage}{\starwidth}
\centering
\begin{tikzpicture}
\input{star-invariants}
\node[draw] at (-\radius*1.2,\radius) {7};
\filldraw [draw=black,fill=red] (v1) circle (\diameter);
\filldraw [draw=black,fill=red] (v2) circle (\diameter);
\filldraw [draw=black,fill=green!40] (v3) circle (\diameter);
\filldraw [draw=black,fill=red] (v4) circle (\diameter);
\filldraw [draw=black,fill=red] (v5) circle (\diameter);
\filldraw [draw=black,fill=white] (v6) circle (\diameter);

\foreach \vertex in {1,...,6}
{
  \node at ({360 - (\vertex - 1) * \angle}: {\radius * 1.0} ) {{\vertex}};
}
\end{tikzpicture}
\end{minipage}

&

\begin{minipage}{\mapwidth}
\begin{flushleft}
$
\begin{array}{@{}r@{\,}c@{\,}l@{}}
1 & \mapsto  & \emptyset \\
2 & \mapsto  & \emptyset \\
3 & \mapsto  & \{ \{ \textcolor{red}{1, 3} \} \} \\
4 & \mapsto  & \emptyset \\
5 & \mapsto  & \{ \{ \textcolor{red}{2, 5} \} \} \\
\end{array}
$
\end{flushleft}
\end{minipage}

&

\begin{minipage}{\starwidth}
\centering
\begin{tikzpicture}
\input{star-invariants}
\node[draw] at (-\radius*1.2,\radius) {8};
\filldraw [draw=black,fill=red] (v1) circle (\diameter);
\filldraw [draw=black,fill=red] (v2) circle (\diameter);
\filldraw [draw=black,fill=green!40] (v3) circle (\diameter);
\filldraw [draw=black,fill=red] (v4) circle (\diameter);
\filldraw [draw=black,fill=green!40] (v5) circle (\diameter);
\filldraw [draw=black,fill=white] (v6) circle (\diameter);

\foreach \vertex in {1,...,6}
{
  \node at ({360 - (\vertex - 1) * \angle}: {\radius * 1.0} ) {{\vertex}};
}
\end{tikzpicture}
\end{minipage}

&

\begin{minipage}{\mapwidth}
\begin{flushleft}
$
\begin{array}{@{}r@{\,}c@{\,}l@{}}
1 & \mapsto  & \emptyset \\
2 & \mapsto  & \emptyset \\
3 & \mapsto  & \{ \{ \textcolor{red}{1, 3} \} \} \\
4 & \mapsto  & \emptyset \\
5 & \mapsto  & \{ \{ \textcolor{red}{2, 5} \} \} \\
\end{array}
$
\end{flushleft}
\end{minipage}

&

\begin{minipage}{\starwidth}
\centering
\begin{tikzpicture}
\input{star-invariants}
\node[draw] at (-\radius*1.2,\radius) {9};
\filldraw [draw=black,fill=red] (v1) circle (\diameter);
\filldraw [draw=black,fill=red] (v2) circle (\diameter);
\filldraw [draw=black,fill=green!40] (v3) circle (\diameter);
\filldraw [draw=black,fill=red] (v4) circle (\diameter);
\filldraw [draw=black,fill=green!40] (v5) circle (\diameter);
\filldraw [draw=black,fill=red] (v6) circle (\diameter);

\foreach \vertex in {1,...,6}
{
  \node at ({360 - (\vertex - 1) * \angle}: {\radius * 1.0} ) {{\vertex}};
}
\end{tikzpicture}
\end{minipage}

&

\begin{minipage}{\mapwidth}
\begin{flushleft}
$
\begin{array}{@{}r@{\,}c@{\,}l@{}}
1 & \mapsto  & \emptyset \\
2 & \mapsto  & \emptyset \\
3 & \mapsto  & \{ \{ \textcolor{red}{1, 3} \} \} \\
4 & \mapsto  & \emptyset \\
5 & \mapsto  & \{ \{ \textcolor{red}{2, 5} \} \} \\
6 & \mapsto  & \{ \{ \textcolor{red}{2, 6} \} \}
\end{array}
$
\end{flushleft}
\end{minipage}

\\[6ex]

\begin{minipage}{\starwidth}
\centering
\begin{tikzpicture}
\input{star-invariants}
\node[draw] at (-\radius*1.2,\radius) {10};
\filldraw [draw=black,fill=red] (v1) circle (\diameter);
\filldraw [draw=black,fill=red] (v2) circle (\diameter);
\filldraw [draw=black,fill=green!40] (v3) circle (\diameter);
\filldraw [draw=black,fill=red] (v4) circle (\diameter);
\filldraw [draw=black,fill=green!40] (v5) circle (\diameter);
\filldraw [draw=black,fill=green!40] (v6) circle (\diameter);

\foreach \vertex in {1,...,6}
{
  \node at ({360 - (\vertex - 1) * \angle}: {\radius * 1.0} ) {{\vertex}};
}
\end{tikzpicture}
\end{minipage}

&

\begin{minipage}{\mapwidth}
\begin{flushleft}
$
\begin{array}{@{}r@{\,}c@{\,}l@{}}
1 & \mapsto  & \emptyset \\
2 & \mapsto  & \emptyset \\
3 & \mapsto  & \{ \{ \textcolor{red}{1, 3} \} \} \\
4 & \mapsto  & \emptyset \\
5 & \mapsto  & \{ \{ \textcolor{red}{2, 5} \} \} \\
6 & \mapsto  & \{ \{ \textcolor{red}{2, 6} \}, \{ \textcolor{green}{3, 6} \} \}
\end{array}
$
\end{flushleft}
\end{minipage}

&

\begin{minipage}{\starwidth}
\centering
\begin{tikzpicture}
\input{star-invariants}
\node[draw] at (-\radius*1.2,\radius) {11};
\filldraw [draw=black,fill=red] (v1) circle (\diameter);
\filldraw [draw=black,fill=red] (v2) circle (\diameter);
\filldraw [draw=black,fill=green!40] (v3) circle (\diameter);
\filldraw [draw=black,fill=white] (v4) circle (\diameter);
\filldraw [draw=black,fill=white] (v5) circle (\diameter);
\filldraw [draw=black,fill=white] (v6) circle (\diameter);

\foreach \vertex in {1,...,6}
{
  \node at ({360 - (\vertex - 1) * \angle}: {\radius * 1.0} ) {{\vertex}};
}
\end{tikzpicture}
\end{minipage}

&

\begin{minipage}{\mapwidth}
\begin{flushleft}
$
\begin{array}{@{}r@{\,}c@{\,}l@{}}
1 & \mapsto  & \emptyset \\
2 & \mapsto  & \emptyset \\
3 & \mapsto  & \{ \{ \textcolor{red}{1, 3} \}, \{ \textcolor{red}{2}, \textcolor{green}{3} \} \} 
\end{array}
$
\end{flushleft}
\end{minipage}

&

\begin{minipage}{\starwidth}
\centering
\begin{tikzpicture}
\input{star-invariants}
\node[draw] at (-\radius*1.2,\radius) {12};
\filldraw [draw=black,fill=red] (v1) circle (\diameter);
\filldraw [draw=black,fill=red] (v2) circle (\diameter);
\filldraw [draw=black,fill=white] (v3) circle (\diameter);
\filldraw [draw=black,fill=white] (v4) circle (\diameter);
\filldraw [draw=black,fill=white] (v5) circle (\diameter);
\filldraw [draw=black,fill=white] (v6) circle (\diameter);

\foreach \vertex in {1,...,6}
{
  \node at ({360 - (\vertex - 1) * \angle}: {\radius * 1.0} ) {{\vertex}};
}
\end{tikzpicture}
\end{minipage}

&

\begin{minipage}{\mapwidth}
\begin{flushleft}
$
\begin{array}{@{}r@{\,}c@{\,}l@{}}
1 & \mapsto  & \emptyset \\
2 & \mapsto  & \{ \{ \textcolor{red}{1, 2} \} \}  
\end{array}
$
\end{flushleft}
\end{minipage}

\\[6ex]

\begin{minipage}{\starwidth}
\centering
\begin{tikzpicture}
\input{star-invariants}
\node[draw] at (-\radius*1.2,\radius) {13};
\filldraw [draw=black,fill=red] (v1) circle (\diameter);
\filldraw [draw=black,fill=green!40] (v2) circle (\diameter);
\filldraw [draw=black,fill=white] (v3) circle (\diameter);
\filldraw [draw=black,fill=white] (v4) circle (\diameter);
\filldraw [draw=black,fill=white] (v5) circle (\diameter);
\filldraw [draw=black,fill=white] (v6) circle (\diameter);

\foreach \vertex in {1,...,6}
{
  \node at ({360 - (\vertex - 1) * \angle}: {\radius * 1.0} ) {{\vertex}};
}
\end{tikzpicture}
\end{minipage}

&

\begin{minipage}{\mapwidth}
\begin{flushleft}
$
\begin{array}{@{}r@{\,}c@{\,}l@{}}
1 & \mapsto  & \emptyset \\
2 & \mapsto  & \{ \{ \textcolor{red}{1, 2} \} \}  
\end{array}
$
\end{flushleft}
\end{minipage}

&

\begin{minipage}{\starwidth}
\centering
\begin{tikzpicture}
\input{star-invariants}
\node[draw] at (-\radius*1.2,\radius) {14};
\filldraw [draw=black,fill=red] (v1) circle (\diameter);
\filldraw [draw=black,fill=green!40] (v2) circle (\diameter);
\filldraw [draw=black,fill=red] (v3) circle (\diameter);
\filldraw [draw=black,fill=white] (v4) circle (\diameter);
\filldraw [draw=black,fill=white] (v5) circle (\diameter);
\filldraw [draw=black,fill=white] (v6) circle (\diameter);

\foreach \vertex in {1,...,6}
{
  \node at ({360 - (\vertex - 1) * \angle}: {\radius * 1.0} ) {{\vertex}};
}
\end{tikzpicture}
\end{minipage}

&

\begin{minipage}{\mapwidth}
\begin{flushleft}
$
\begin{array}{@{}r@{\,}c@{\,}l@{}}
1 & \mapsto  & \emptyset \\
2 & \mapsto  & \{ \{ \textcolor{red}{1, 2} \} \}  \\
3 & \mapsto & \{ \{ \textcolor{red}{1, 3} \} \}
\end{array}
$
\end{flushleft}
\end{minipage}

&

\begin{minipage}{\starwidth}
\centering
\begin{tikzpicture}
\input{star-invariants}
\node[draw] at (-\radius*1.2,\radius) {15};
\filldraw [draw=black,fill=red] (v1) circle (\diameter);
\filldraw [draw=black,fill=green!40] (v2) circle (\diameter);
\filldraw [draw=black,fill=green!40] (v3) circle (\diameter);
\filldraw [draw=black,fill=white] (v4) circle (\diameter);
\filldraw [draw=black,fill=white] (v5) circle (\diameter);
\filldraw [draw=black,fill=white] (v6) circle (\diameter);

\foreach \vertex in {1,...,6}
{
  \node at ({360 - (\vertex - 1) * \angle}: {\radius * 1.0} ) {{\vertex}};
}
\end{tikzpicture}
\end{minipage}

&

\begin{minipage}{\mapwidth}
\begin{flushleft}
$
\begin{array}{@{}r@{\,}c@{\,}l@{}}
1 & \mapsto  & \emptyset \\
2 & \mapsto  & \{ \{ \textcolor{red}{1, 2} \} \}  \\
3 & \mapsto & \{ \{ \textcolor{red}{1, 3} \} \}
\end{array}
$
\end{flushleft}
\end{minipage}

\\[6ex]

\begin{minipage}{\starwidth}
\centering
\begin{tikzpicture}
\input{star-invariants}
\node[draw] at (-\radius*1.2,\radius) {16};
\filldraw [draw=black,fill=red] (v1) circle (\diameter);
\filldraw [draw=black,fill=green!40] (v2) circle (\diameter);
\filldraw [draw=black,fill=green!40] (v3) circle (\diameter);
\filldraw [draw=black,fill=red] (v4) circle (\diameter);
\filldraw [draw=black,fill=white] (v5) circle (\diameter);
\filldraw [draw=black,fill=white] (v6) circle (\diameter);

\foreach \vertex in {1,...,6}
{
  \node at ({360 - (\vertex - 1) * \angle}: {\radius * 1.0} ) {{\vertex}};
}
\end{tikzpicture}
\end{minipage}

&

\begin{minipage}{\mapwidth}
\begin{flushleft}
$
\begin{array}{@{}r@{\,}c@{\,}l@{}}
1 & \mapsto  & \emptyset \\
2 & \mapsto  & \{ \{ \textcolor{red}{1, 2} \} \}  \\
3 & \mapsto & \{ \{ \textcolor{red}{1, 3} \} \} \\
4 & \mapsto & \emptyset 
\end{array}
$
\end{flushleft}
\end{minipage}

&

\begin{minipage}{\starwidth}
\centering
\begin{tikzpicture}
\input{star-invariants}
\node[draw] at (-\radius*1.2,\radius) {17};
\filldraw [draw=black,fill=red] (v1) circle (\diameter);
\filldraw [draw=black,fill=green!40] (v2) circle (\diameter);
\filldraw [draw=black,fill=green!40] (v3) circle (\diameter);
\filldraw [draw=black,fill=red] (v4) circle (\diameter);
\filldraw [draw=black,fill=red] (v5) circle (\diameter);
\filldraw [draw=black,fill=white] (v6) circle (\diameter);

\foreach \vertex in {1,...,6}
{
  \node at ({360 - (\vertex - 1) * \angle}: {\radius * 1.0} ) {{\vertex}};
}
\end{tikzpicture}
\end{minipage}

&

\begin{minipage}{\mapwidth}
\begin{flushleft}
$
\begin{array}{@{}r@{\,}c@{\,}l@{}}
1 & \mapsto  & \emptyset \\
2 & \mapsto  & \{ \{ \textcolor{red}{1, 2} \} \}  \\
3 & \mapsto & \{ \{ \textcolor{red}{1, 3} \} \} \\
4 & \mapsto & \emptyset \\
5 & \mapsto & \emptyset 
\end{array}
$
\end{flushleft}
\end{minipage}

&

\begin{minipage}{\starwidth}
\centering
\begin{tikzpicture}
\input{star-invariants}
\node[draw] at (-\radius*1.2,\radius) {18};
\filldraw [draw=black,fill=red] (v1) circle (\diameter);
\filldraw [draw=black,fill=green!40] (v2) circle (\diameter);
\filldraw [draw=black,fill=green!40] (v3) circle (\diameter);
\filldraw [draw=black,fill=red] (v4) circle (\diameter);
\filldraw [draw=black,fill=red] (v5) circle (\diameter);
\filldraw [draw=black,fill=red] (v6) circle (\diameter);

\foreach \vertex in {1,...,6}
{
  \node at ({360 - (\vertex - 1) * \angle}: {\radius * 1.0} ) {{\vertex}};
}
\end{tikzpicture}
\end{minipage}

&

\begin{minipage}{\mapwidth}
\begin{flushleft}
$
\begin{array}{@{}r@{\,}c@{\,}l@{}}
1 & \mapsto  & \emptyset \\
2 & \mapsto  & \{ \{ \textcolor{red}{1, 2} \} \}  \\
3 & \mapsto & \{ \{ \textcolor{red}{1, 3} \} \} \\
4 & \mapsto & \emptyset \\
5 & \mapsto & \emptyset \\
6 & \mapsto & \emptyset 
\end{array}
$
\end{flushleft}
\end{minipage}

\end{tabular}

\end{center}
\caption{Graph Colouring with Backjumping}\label{fig-animate}
\end{figure}

Green has yet to be tried for vertex~3, hence backtracking is applied to
undo the assignment at vertex~3 and reassign it green (diagram 5).
The conflict $\{ 1, 3 \}$ is retained: the assignment to vertex 1 is still red
but vertex~3 is now green. Thus, both vertices of the conflict, with the possible exception of the last,
preserve their initial colours, which is a general pattern.

The next conflict arises at vertex~5 (diagram 7), causing the
conflict $\{ 2, 5 \}$ to be recorded. Not all colours have been considered at vertex~5 so
again backtracking undoes the assignment to vertex~5 and it is reassigned green.  
Next, vertex~6 is coloured red, which conflicts with vertex~2 
(diagram 9).  Again, not all colours have been tried
at vertex~6, thus the vertex is reassigned to green, which then conflicts with vertex~3 
(diagram 10).  Notice that the first vertices (in the vertex ordering)
of $\{ 2, 5 \}$, $\{ 2, 6 \}$ and $\{ 3, 6 \}$ retain their initial colour.

\subsection{Conflict Analysis for Graph Colouring}

Now that all colours have been tried at vertex~6, backjumping is deployed after
a form of conflict analysis which infers the target of the backjump.  The
conflicts for vertex~6 are $\{ 2, 6 \}$ and $\{ 3, 6 \}$, indicating that the
conflicts at vertex 6 involve the assignments of vertices 2, 3 and 6, but not
vertices 4 and 5.  Moreover, one conflict occurs when vertices 2~and~6 are both
red, and the other occurs when vertices 3~and~6 are both green.  Hence, $\{ 2, 3 \}$
is also a conflict, where 2 is red and 3 is green, since this partial
assignment is incompatible with the edge constraints, irrespective of the
colour assigned to vertex~6. Therefore, a solution cannot be found without either
reassigning the colour at vertex~2, or vertex~3, or both. The conflict 
set of vertex~3
is augmented with 
$\{ 2, 3 \}$
to give 
$\{ \{ 1, 3 \}, \{ 2, 3 \} \}$ (diagram 11).  
Since vertex~3 was assigned
more recently than vertex~2, it is selected as the target of the backjump and
search resumes at vertex~3.

It should be noted that although the vertices of each conflict are coloured in Fig.~\ref{fig-animate}, it is not necessary
to introduce additional colour assignments, one per conflict, to record these colours. To see this, observe
how the colours of 2, 3 in $\{ 2, 3 \}$ align with
the current colour assignment because the 2 of $\{ 2, 6 \}$ and the 3 of $\{ 3, 6 \}$
also match with the current colour assignment.   Thus, the
first vertices of both $\{ 1, 3 \}$ and $\{ 2, 3 \}$ match the current
colour assignment.  Hence, the colours of the
vertices of a conflict, with the possible exception of the last, match those of the current colour assignment, a property which holds inductively. 

Vertex~3 has already been assigned to both red and green, so again a form of conflict
analysis is applied to infer that the partial assignment on the
vertices 1 and 2 cannot be extended to a complete solution, whatever
the colour of vertex~3.
Therefore, $\{ 1, 2 \}$ is also a conflict, which is associated with vertex~2.
Note how the colours of $\{ 1, 2 \}$, which are both red, match the current colour assignment.
This conflict cannot be remedied without reassigning either vertex~1 or vertex~2 or both.  The higher of the two, vertex~2, is thus taken as the target of the backjump,
which is then coloured green (diagram 13).
Conflicts on vertex~3 result in
reassignment of that vertex to green (diagram 15), 
before the search proceeds to find a
complete satisfying colour assignment (diagram 18).

Finally, observe that the map data-structure can be simplified by replacing
each set of conflicts with a single set which is the union of all its conflicts \cite{bruynooghe04enchancing}.  
Thus, the map
for diagrams 9, 10, 11 and 12, are replaced by:
\[
\begin{array}{@{}r@{\,}c@{\,}l@{}}
1 & \mapsto  & \emptyset \\
2 & \mapsto  & \emptyset \\
3 & \mapsto  & \{ 1, 3 \} \\
4 & \mapsto  & \emptyset \\
5 & \mapsto  & \{ 2, 5 \} \\
6 & \mapsto  & \{ 2, 6 \}
\end{array}
\qquad
\begin{array}{@{}r@{\,}c@{\,}l@{}}
1 & \mapsto  & \emptyset \\
2 & \mapsto  & \emptyset \\
3 & \mapsto  & \{ 1, 3 \} \\
4 & \mapsto  & \emptyset \\
5 & \mapsto  & \{ 2, 5 \} \\
6 & \mapsto  & \{ 2, 3, 6 \}
\end{array}
\qquad 
\begin{array}{@{}r@{\,}c@{\,}l@{}}
1 & \mapsto  & \emptyset \\
2 & \mapsto  & \emptyset \\
3 & \mapsto  & \{ 1, 2, 3 \} 
\end{array}
\qquad
\begin{array}{@{}r@{\,}c@{\,}l@{}}
1 & \mapsto  & \emptyset \\
2 & \mapsto  & \{ 1, 2 \} 
\end{array}
\]
Observe how entry
$3 \mapsto  \{ 1, 2, 3 \} $ of the third map given immediately above
can be found by unioning the vertex sets for
$3 \mapsto  \{ 1, 3 \}$
and
$6 \mapsto  \{ 2, 3, 6 \}$ of its predecessor
and then eliminating vertex~6, which is the vertex whose colouring induced the conflict analysis.
In fact, not all
the map needs to be accessed simultaneously:  only the set of vertices for the highest
identifier. This allows the map to be organised as a stack of lists, where only the topmost
list of vertices on the stack is modified at any one time.
This gives a straightforward data-structure for implementing conflict analysis. 

Notice that it is possible that an assignment leads to several conflicts.  
Here, the standard approach is taken -- one conflict is selected to inform 
the backjump, and it is possible that one of the other conflicts is then
encountered as search continues.  An alternative approach would be to record
all conflicts, and backjump to the shallowest point in the search tree to
guarantee that none of these would be encountered again.
In the Prolog implementation considered
in the next section, if there are several conflicts, 
the scheduler will determine which conflict is first
encountered and used to make the backjump.  The correctness of backjumping
is addressed in \cite{kondrak97theoretical}.

\subsection{Backjumping in Prolog}

\newcommand{\guillotine}{\ensuremath{\mid}\,}
\newcommand{\notequal}{\ensuremath{\setminus \text{=} \;\;}}
\newcommand{\notunify}{\ensuremath{\setminus \text{==} \;\;}}

\begin{figure}[t!]
\tt
\begin{tabbing}
4ex\=4ex\=4ex\=4ex\=4ex\=4ex\=4ex\kill
colour(Vars, Values, Cs) :- \\
\> problem\_setup(Vars, Values, Cs, Pairs, AllIds), \\
\> search(Pairs, [], AllIds). \\
 \\
problem\_setup(Vars, Values, Cs, Pairs, AllIds) :-  \\ 
\> pairs(Vars, 1, Values, Pairs, AllIds),  \\
\> setup\_checks(Cs). \\
\\
pairs([], \_Id, \_Values, [], []). \\
pairs([Var | Vars], Id, Values, [Var-Values | Pairs], [Id | Ids]) :-  \\
\> Var = \_-Id, \\
\> NextId is Id + 1, \\
\> pairs(Vars, NextId, Values, Pairs, Ids). \\
 \\
setup\_checks([]). \\
setup\_checks([C | Cs]) :- \\
\> post(C), \\
\> setup\_checks(Cs). \\
 \\
post(X-XId \notequal Y-YId) :- \\
\> suspend(X, Y, XId, YId). \\
 \\
:- block suspend(-, ?, ?, ?), suspend(?, -, ?, ?). \\
suspend(X, Y, XId, YId) :- \\
\> ( \\
\> \>inconsistent(X, Y) \\
\> -> \\
\> \> MaxId is max(XId, YId), \\
\> \> MinId is min(XId, YId), \\
\> \> throw(ball(MaxId, [MinId, MaxId]))  \\
\> ; \\
\> \> true \\
\> ). \\
 \\
inconsistent(X, Y) :- X == Y.
\end{tabbing}
\caption{Setting up the checks for Graph Colouring}\label{fig-colour-constraints}
\end{figure}

Colouring can be realised by adopting a test-and-generate model in which
checks suspend on their variables until they become sufficiently
instantiated to apply the test.  These checks then coroutine with a generator phase which binds
the variables, one by one, that represent the colours of the vertices. 
SICStus code is provided to achieve this, where
Fig.~\ref{fig-colour-constraints}
sets up the checks
which define the colouring problem and Fig.~\ref{fig-colour-backjump} 
applies labelling, with backjumping, to search for a satisfying assignment.

\paragraph{Setting up the checks for colouring} 

The predicate \verb|colour(Vars, Values, Cs)| solves a
colouring problem, where \verb|Vars| is a list that specifies
the colour which is assigned to each vertex,
\verb|Values| defines the range of colours
that can be assigned at each vertex, and
\verb|Cs| is a list of disequalities specifying the edge constraints. 
For the problem instance in Fig.~\ref{fig-animate} the initial call would have:
\begin{verbatim}
Vars = [One, Two, Three, Four, Five, Six]
Values = [red, green]
Cs = [One \= Three, Two \= Five, Two \= Six, 
                    Three \= Six, Three \= Four]
\end{verbatim}

The initial call first invokes \verb|problem_setup(Vars, Values, Cs, Pairs, AllIds)|
that is responsible for setting up \verb|Vars|, \verb|Pairs| and \verb|AllIds|,
and has as its second goal, \verb|search(Pairs, [], AllIds)| for controlling the search.
In setting up the problem, the 
predicate \verb|pairs(Vars, 1, Values, Pairs, AllIds)| 
instantiates each element of the \verb|Pairs| list to a
term \verb|Var-Values| where \verb|Var| is itself
instantiated to a pair \verb|Value-Id| such that \verb|Value| is drawn from the list \verb|Values|
and
\verb|Id| is a (ground) identifier; \verb|AllIds| is instantiated to a list of all identifiers.
The identifiers are numeric, indicating the position
of \verb|Var-Values| within \verb|Pairs|, which tallies with the order in which variables are (later) assigned.  
The predicate \verb|setup_checks(Cs)|
posts each check given in a list of disequalities \verb|Cs|.
Each goal \verb|post(X-XId \= Y-YId)|
invokes \verb|suspend(X, Y, XId, YId)| whose
block declaration
specifies that the \verb|suspend| goal should not be called
until both \verb|X| and \verb|Y| are instantiated. 
If inconsistency is detected on instantiation then
the term \verb|ball(MaxId, CIds)| is thrown
where \verb|MaxId| is the largest identifier
of the set \verb|CIds| formed from 
\verb|XId| and \verb|YId| which identify the variables
involved in the conflict.
Note that coroutining is described here
using SICStus's block declarations \cite{carlsson12sicstus}. Although coroutining is not
part of the ISO standard, other mainstream Prolog systems
provide similar control constructs for delaying goals, such as \verb|when| or \verb|freeze|.

\begin{figure}[t!]
\tt
\begin{tabbing}
4ex\=4ex\=4ex\=4ex\=4ex\=4ex\=4ex\kill

search([], \_, AllIds) :- \\
\> succeed(AllIds). \\
search([(Var-Id)-[Value | RestValues] | Pairs], ConflictIds, AllIds) :- \\
\> catch( \\
\> \> ( \\
\> \> \> bind(Var,Value), \\
\> \> \> search(Pairs, [], AllIds) \\
\> \> ), \\
\> \> ball(Id, CIds), \\
\> \> ( \\
\> \> \> update\_conflict(RestValues, CIds, ConflictIds, Id, NewConflictIds), \\
\> \> \> search([(Var-Id)-RestValues | Pairs], NewConflictIds, AllIds) \\
\> \> ) \\
\> ). \\
 \\
bind(Var, Value) :- Var = Value. \\
 \\
update\_conflict([], CIds, ConflictIds, Id, NewConflictIds) :- \\
\> merge(CIds, ConflictIds, NewConflictIds), \\
\> delete(NewConflictIds, Id, RestConflictIds), \\
\> max\_member(MaxId, RestConflictIds), \\
\> throw(ball(MaxId, RestConflictIds)). \\
update\_conflict([\_Value | \_RestValues], CIds, ConflictIds, \_Id, NewConflictIds) :- \\
\> merge(CIds, ConflictIds, NewConflictIds). \\
 \\
succeed(\_). \\
succeed(AllIds) :- \\
\> max\_member(MaxId, AllIds), \\
\> throw(ball(MaxId, AllIds)). \\
\end{tabbing}
\caption{Backjumping search for Graph Colouring}\label{fig-colour-backjump}
\end{figure}

\paragraph{Backjumping for colouring} 
The predicate \verb|search(Pairs, ConflictIds, AllIds)| assigns
the variables of \verb|Pairs| to satisfy the constraints, failing if there is no solution.
The \verb|ConflictIds| argument
maintains a conflict set for the 
variable of the first pair of \verb|Pairs|.

Consider first the second clause of \verb|search| which is responsible
for orchestrating backjumping.
Each meta-call \verb|catch(Goal, Catcher, RecoveryGoal)|
has \verb|Goal| that is concerned with
assigning one variable, identified by \verb|Id|, to the colour \verb|Value|.
Binding \verb|Var| to \verb|Value| might wake up blocked calls to \verb|suspend|,
which will lead to consistency checks.
If binding \verb|Var| to \verb|Value| does not lead to inconsistency
then search proceeds to search for an assignment to
the remaining \verb|Pairs|.
If inconsistency is discovered 
\verb|ball(MaxId, CIds)| is thrown by a check in \verb|suspend|
and the call stack is unwound to
the first enclosing \verb|catch| 
for which \verb|Catcher| unifies with \verb|ball(MaxId, CIds)|.
The \verb|Catcher| term of
each meta-call is \verb|ball(Id, CIds)| where \verb|Id| is ground and \verb|CIds| is a variable,
thus any \verb|catch| which
intercepts \verb|ball(MaxId, CIds)| must possess an identifier \verb|Id| 
which matches \verb|MaxId|. 
In fact, 
this realises backtracking; 
if a call to \verb|bind| assigns \verb|Var| (with identifier \verb|Id|),
wakes \verb|suspend| in Fig~\ref{fig-colour-constraints} and inconsistency
is discovered, then both variables are instantiated, hence \verb|MaxId| must
be \verb|Id| and the \verb|catch| directly enclosing the call to \verb|bind|
handles the exception.  This then 
allows the conflict information
in \verb|CIds| to be passed back to the point where search resumes.  
An earlier ancestor will only intercept a
ball thrown in \verb|update_conflict| which is realising backjumping.
Note that \verb|bind(Var, Value)| can unblock several \verb|suspend| goals. Yet when the first goal resumes
it will throw its ball, undoing the binding, so that the other \verb|suspend| goals become blocked again.

\verb|RecoveryGoal| has two calls, the first to \verb|update_conflict| maintains conflict information for
backjumping, and the second continues search.
If there are further colours to 
be tried, the second clause
of \verb|update_conflict| merges the conflict information for the current failure with that for 
previous failures, without duplication.  The call to \verb|search| will then assign the next colour from
\verb|RestValues|.  If there are no further colours to be tried, that is \verb|RestValues| is empty, 
backjumping should occur.  The first clause of \verb|update_conflict| enables this.  
Conflict information is merged, the current assignment identifier is removed from the conflict list, 
and then the highest identifier remaining is the backjump level, hence this and the conflict information is 
thrown as \verb|ball(MaxId, RestConflictIds)|.  This achieves backjumping by 
unwinding the call stack to where the
variable with identifier \verb|MaxId| is bound, whilst communicating
the new conflict \verb|RestConflictIds| to that part of the search.
At this point, either a colour
remains to be assigned or backjumping is again applied, and so search continues.
Observe that undoing and then reassigning a variable to another colour, which is the
essence of backtracking, is realised entirely using \verb|catch| and \verb|throw|.

If the first clause of \verb|search| is matched then all variables will
have been assigned a colour.  This clause of \verb|search| 
invokes the goal \verb|succeed(AllIds)|, which will immediately succeed thereby
returning the solution to the colouring problem. If another answer is requested
then a throw is used to reactivate search, the \verb|MaxId| selected
from  \verb|AllIds| which is list of all the variable identifiers. 
This results in search backtracking into the non-conflicting solution.
The call to \verb|succeed(AllIds)| can be
omitted if it is sufficient to compute a single answer.  
It should be noted that all control is provided by catch and throw: it is not necessary
to resort to a mutable database to maintain the
conflicts and direct search \cite{bruynooghe04enchancing}.

Search as presented in Fig.~\ref{fig-colour-constraints} and \ref{fig-colour-backjump} 
provides a template for implementing backjumping. 
Predicate \verb|search| wraps \verb|catch(Goal, Catcher, RecoveryGoal)| where the role of 
\verb|Goal| is to
bind variables to values, 
but if conflicts arise they are 
described by the term
\verb|ball(Id, CIds)| and 
caught by \verb|Catcher|, 
and then the role of \verb|RecoveryGoal| is to update conflict information and either continue search, 
or backjump, as appropriate.


\section{SAT}\label{sec:sat}

Figure~\ref{fig:sat-solver} lists the code for a Prolog SAT solver, adapted
from~\cite{howe10pearl,howe12pearl}, that uses watched literals to realise unit
propagation.  Given a propositional formula $f$ in CNF over a set of variables
$X$, and a partial (truth) function 
$\theta : X \to \{ \mathit{true}, \mathit{false} \}$, 
unit propagation examines each clause of $f$  to deduce another partial
function $\theta' : X \to \{ \mathit{true}, \mathit{false} \}$ that extends
$\theta$ and 
that, if $\theta$ can be extended to satisfy $f$,
necessarily holds.
For example, suppose
$X = \{x, y, z, u, v, w\}$, 
$f = (\neg x \vee z \vee \neg y) \wedge (\neg z \vee \neg u) \wedge (u \vee w \vee \neg v) \wedge (\neg w \vee v)$
and
$\theta = \{ x \mapsto \mathit{true}, y \mapsto \mathit{true} \}$.
In this instance the clause $(\neg x \vee z \vee \neg y)$ is unit, because it has only one
unbound variable, $z$. 
Therefore, it can be deduced that, given $\theta$, for the clause to be
satisfied $z \mapsto \mathit{true}$. Moreover, for
$(\neg z \vee \neg u)$ to be satisfied, it follows that
$u \mapsto \mathit{false}$.  
The satisfaction of the remaining two clauses
depends on two
unknowns, $v$ and $w$, hence no further information can be deduced from them.
Therefore, $\theta' = \theta \cup \{  z \mapsto \mathit{true}, u \mapsto \mathit{false} \}$.

\begin{figure}[t!]
\tt
\begin{tabbing}
4ex\=4ex\=4ex\=4ex\=4ex\=4ex\=4ex\kill
sat(Clauses, Vars) :- \\
\> watch\_clauses(Clauses), \\
\> search(Vars). \\
\\
search([]). \\
search([Var \guillotine\ Vars]) :- \\
\> (Var = true; Var = false), \\
\> search(Vars). \\
\\
watch\_clauses([]). \\
watch\_clauses([Clause \guillotine\ Clauses]) :- \\
\> watch\_clause(Clause), \\
\> watch\_clauses(Clauses). \\
\\
watch\_clause([Pol-Var \guillotine\ Pairs]) :- set\_watch(Pairs, Var, Pol). \\
\\
set\_watch([], Var, Pol) :- Var = Pol. \\
set\_watch([Pol2-Var2 \guillotine\ Pairs], Var1, Pol1) :- \\
\> watch(Var1, Pol1, Var2, Pol2, Pairs). \\
\\
:- block watch(-, ?, -, ?, ?). \\
watch(Var1, Pol1, Var2, Pol2, Pairs) :- \\
\> ( \\
\> \> nonvar(Var1) \\
\> -> \\
\> \> update\_watch(Var1, Pol1, Var2, Pol2, Pairs) \\
\> ; \\
\> \> update\_watch(Var2, Pol2, Var1, Pol1, Pairs) \\
\> ). \\
\\
update\_watch(Var1, Pol1, Var2, Pol2, Pairs) :- \\
\> ( \\
\> \> Var1 == Pol1 \\
\> -> \\
\> \> true \\
\> ; \\
\> \> set\_watch(Pairs, Var2, Pol2) \\
\> ). \\ 
\end{tabbing}
\caption{A (vanilla) SAT solver using watched literals \protect\cite{howe10pearl,howe12pearl}}
\label{fig:sat-solver}
\end{figure}

Searching for a satisfying assignment of $f$ 
proceeds as follows: starting from an
empty truth function $\theta = \emptyset$, unit propagation is applied to
$\theta$ until either no further propagation is possible or a contradiction is
established.  In the first case, if all clauses are satisfied then $f$ is
satisfied, else an unassigned variable occurring in $f$, for instance $x$, is selected
and the assignment $x\mapsto \mathit{true}$ is added to $\theta$.  In the
second case,  search backtracks to a previous assignment, $y \mapsto
\mathit{true}$ say, then adds $y\mapsto \mathit{false}$ to $\theta$ and
continues with unit propagation.

The watched literals technique is founded on the simple observation that a
particular clause is unit if it does not contain two unassigned
variables~\cite{moskewicz01chaff}. Therefore, for each clause of a problem, two
unassigned variables are watched; propagation may occur once either is assigned. 
It is not enough to watch just one variable because this is, in general,
not sufficient for detecting if a clause becomes unit: the watch might be on the other, unassigned variable.
The SAT solver in Fig.~\ref{fig:sat-solver} takes a problem in CNF,
specified as a list of clauses, and a list of variables. Each clause is itself
a list of pairs \verb|Pol-Var|, where \verb|Var| is a propositional variable,
and \verb|Pol| indicates whether the variable has positive or negative polarity
by being either \verb|true| or \verb|false| respectively. 
For the introductory example above, the initial call to \verb|sat/2| would have:
\begin{verbatim}
Clauses = [[false-X, true-Z, false-Y], [false-Z, false-U],
           [true-U, true-W, false-V], [false-W, true-V]]
Vars = [X, Y, Z, U, V, W]
\end{verbatim}
For each clause, the
\verb|watch_clause| predicate invokes \verb|set_watch| that, in
turn, selects the first two variables in the clause to be watched.
In the above, for the first clause this leads to the invocation of 
\verb|watch(X, false, Z, true, [false-Y])|; with neither \verb|X| nor 
\verb|Z| instantiated this goal suspends via a delay declaration 
(specified using \verb|block|
in SICStus syntax).
When a variable is instantiated \verb|watch| resumes and executes
\verb|update_watch|. If the instantiated variable matches its polarity, the
clause is satisfied, and \verb|update_watch| exits successfully,
otherwise another variable is selected for watching.  For the clause
being considered, if \verb|X| is bound to \verb|true|, then \verb|set_watch([false-Y], Z, true)|
will be called which in turn will lead to the suspended
\verb|watch(Z, true, Y, false, [])|.
If only one unbound variable
remains, \verb|set_watch| realises unit propagation and assigns
that variable so that the clause is satisfied.  
So if \verb|Y| is bound to \verb|true|, after
waking \verb|watch|, 
\verb|set_watch| instantiates \verb|Z| to be \verb|true| by
unit propagation; further, the second clause leads to \verb|U| 
being instantiated to \verb|false|. 
If there are no unassigned variables, or assigned and satisfying
variables, then the clause is unsatisfiable and the \verb|set_watch|  goal will
fail, and search will backtrack.  If the example being considered is further 
extended by binding \verb|V| to \verb|true|, 
then the third clause will infer by unit propagation that \verb|W| is
instantiated to \verb|true|.  The fourth clause, which was suspended as \verb|watch(W, false, V, true, [])|, will wake and lead to the call \verb|set_watch([], true, false)|, 
which fails.
The search is realised using the \verb|search| predicate to assign
variables to \verb|true| or \verb|false|. The search for a
satisfying assignment then proceeds as previously described, simply through 
Prolog backtracking.


\begin{figure}[t!]
\tt
\begin{tabbing}
4ex\=4ex\=4ex\=4ex\=4ex\=4ex\=4ex\kill
sat\_cdcl(Clauses, Vars) :-\\
\>	put\_learnt([]),\\
\>	ids(Vars, Ids, IdMap),\\
\>	vars(Vars, Ids),\\
\>	watch\_clauses(Clauses),	\\
\>      search\_setup(Vars, IdMap, 1).\\
\\
ids([Var | Vars], Ids, IdMap) :- \\
\>	length(Vars, Length), \\
\>	numlist(0, Length, Ids), \\
\>	maplist(zip, Ids, [Var | Vars], IdsVars), \\ 
\>	list\_to\_assoc(IdsVars, IdMap).\\
\\
zip(X, Y, X-Y).\\
\\
vars([], []).\\
vars([\_-imp(\_Level, Id, \_Pol, \_SubWhys) | Vars], [Id | Ids]) :- \\
\>	vars(Vars, Ids). \\
\end{tabbing}
\caption{Initial call and set up}
\label{fig:cdclSetup}
\end{figure}

\begin{figure}[t!]
\tt
\begin{tabbing}
4ex\=4ex\=4ex\=4ex\=4ex\=4ex\=4ex\kill
watch\_clause([Pol-VarWhy \guillotine\ Pairs]) :- set\_watch(Pairs, VarWhy, Pol, []). \\
\\
set\_watch([], Var-Why, Pol, Whys) :- \\
\> ( \\
\> \> (nonvar(Var), Var \notunify Pol) \\
\> -> \\
\> \> conflict([Why | Whys]) \\
\> ; \\
\> \> unit(Var-Why, Pol, Whys) \\
\> ). \\
set\_watch([Pol2-(Var2-Why2) \guillotine\ Pairs], Var1-Why1, Pol1, Whys) :- \\
\> watch(Var1, Why1, Pol1, Var2, Why2, Pol2, Pairs, Whys). \\
\\
:- block watch(-, ?, ?, -, ?, ?, ?, ?). \\
watch(Var1, Why1, Pol1, Var2, Why2, Pol2, Pairs, Whys) :- \\
\> ( \\
\> \> nonvar(Var1) \\
\> -> \\
\> \> update\_watch(Var1, Why1, Pol1, Var2, Why2, Pol2, Pairs, Whys) \\
\> ; \\
\> \> update\_watch(Var2, Why2, Pol2, Var1, Why1, Pol1, Pairs, Whys) \\
\> ). \\
\\
update\_watch(Var1, Why1, Pol1, Var2, Why2, Pol2, Pairs, Whys) :- \\
\> ( \\
\> \> Var1 == Pol1 \\
\> -> \\
\> \> true \\
\> ; \\
\> \> set\_watch(Pairs, Var2-Why2, Pol2, [Why1 \guillotine\ Whys]) \\
\> ). \\
\\
unit(Var-Why, Pol, Whys) :- \\
\> ( \\
\> \> nonvar(Var) \\
\> -> \\
\> \> true \\
\> ; \\
\> \> max\_member(imp(MaxLevel, \_, \_, \_), [imp(0-0, \_, \_, \_) \guillotine\ Whys]), \\
\> \> increment\_sublevel(MaxLevel, NextLevel),\\
\> \> Why = imp(NextLevel, \_Id, Pol, Whys), \\
\> \> Var = Pol \\
\> ). \\
\\
increment\_sublevel(Level-SubLvl, Level-SubLvl1) :- SubLvl1 is SubLvl + 1. \\
\end{tabbing}
\caption{Setting up the propagators for SAT}
\label{fig:difSetup}
\end{figure}

\subsection{Backjumping in a Prolog SAT Solver}

The goal of this paper is to augment the Prolog SAT solver in Fig.~\ref{fig:sat-solver} with CDCL.   
This is achieved by following the backjumping template from the previous
section, where alongside the backjump, learnt clauses are added to the problem
description.   This section describes how backjumping for SAT is built, including 
the subtleties arising from accommodating clause learning.  Figures~\ref{fig:cdclSetup}, \ref{fig:difSetup},  \ref{fig:backjump}, \ref{fig:conflict} and \ref{fig:reinstate}
extend the SAT solver of Fig.~\ref{fig:sat-solver}
with infrastructure for backjumping and conflict analysis.

\paragraph{Variables and Implication Graphs} 
In Fig.~\ref{fig:cdclSetup}, the initial call to the CDCL solver is given, along with the set up of
the variables. 
An identifier map, \verb|IdMap|,  which associates a ground identifier with the variable it represents, 
is created during set up to later reconstruct clauses during learning.  This mapping is constructed
up-front prior to invoking \verb|watch_clauses|.
An implication graph \cite{marquessilva09cdcl} for a variable \verb|Var| is conceptually a DAG in
which each node is represented as a term
\verb|imp(Level, Id, Value, Whys)| where \verb|Level| records the decision
level at which \verb|Var| was assigned; \verb|Id| is a (ground) identifier
that is unique to the variable (used solely for deriving a ground
representation of a learnt clause); \verb|Value| is the truth value bound to
the variable; and \verb|Whys| is itself a list of implication graphs,
interpreted as subtrees.  
Each
propositional variable of the vanilla solver is replaced with a compound term
\verb|Var-Why| that pairs the variable \verb|Var| with an implication graph
\verb|Why| that explains its instantiation.  
At the set up stage, the \verb|Why|
term of each pair \verb|Var-Why| is unified with \verb|imp(_Level, Id, _Value, _Whys)| where
\verb|Id| is the identifier for \verb|Var| that ensures that each implication graph always carries its identifier.

Recall that in colouring, the decision level 
at which a variable is instantiated matches the position of the variable in
the list \verb|Vars| which also corresponds to its identifier. A consequence of
unit propagation, however, is that the instantiation order does
not necessarily follow the ordering of \verb|Vars|, hence the decision level
does not necessarily match the identifier.  Therefore, 
 \verb|Level| and \verb|Id| are separately recorded in the \verb|imp| structure.
Furthermore, 
a learning solver reasons about the order
in which variables are instantiated. Since, even at the same decision level, the instantiation of one variable
can trigger the instantiation of another, the first element of \verb|imp| is actually a pair \verb|Level-SubLevel|
where the integer \verb|SubLevel| records the order
in which a variable is instantiated within a given \verb|Level|.

\paragraph{Setting up propagators for SAT}
Figure~\ref{fig:difSetup} presents an enhanced version
of \verb|watch_clause| that supports conflict analysis.
Unit propagation is extended to record the reason why a
propositional variable is bound to a particular truth value.  
Since the \verb|Var| argument of  \verb|watch_clause| is
replaced by a pair \verb|Var-Why|, the two \verb|Var1| and
\verb|Var2| arguments of \verb|watch| are accompanied with two additional arguments
\verb|Why1| and \verb|Why2|.  Then, the goal \verb|conflict| uses the
\verb|Why| for \verb|Var| and a list of implication graphs, \verb|Whys|,
for the other variables of the clause to diagnose the cause of the
conflict, which is summarised as a clause.  The goal ultimately
terminates by throwing a ball which includes the learnt clause, or
triggers failure when an inconsistency is found with the clauses posted during set up.
Notice how \verb|Whys| is accumulated as each clause is traversed in \verb|update_watch|.

The \verb|unit(Var-Why, Pol, Whys)| goal calculates the decision
level for a variable as part of the construction
of its implication graph \verb|Why|.  This predicate
is only invoked from \verb|set_watch| when
\verb|Var| is either uninstantiated or bound to \verb|Pol|.
The latter case is vacuous, but the former case of \verb|unit| 
binds \verb|Var| to the truth value \verb|Pol| and creates
\verb|Why| which records the reason for the binding.  The \verb|max_member| predicate 
harvests the maximum of the levels of implication graphs \verb|Whys|
of the other variables collected as the clause is traversed.
The additional \verb|imp(0-0, _, _, _)| term ensures that \verb|MaxLevel|
is well-defined even when inconsistency is detected
prior to any assignment by \verb|search|.  The predicate \verb|increment_sublevel|
merely increments the sublevel.
The unification \verb|Var = Pol| follows the bind to \verb|Why|
to ensure that every instantiated propositional
variable is associated with a complete implication graph.

\begin{figure}[t!]
\tt
\begin{tabbing}
4ex\=4ex\=4ex\=4ex\=4ex\=4ex\=4ex\kill
search\_setup(Vars, IdMap, Level) :- \\
\>        catch(\\
\>\>                 search(Vars, IdMap, Level),\\
\>\>                 ball(0, Clause),\\
\>\>                 (\\
\>\>\>                     update\_conflict(1, Clause, 0, IdMap),\\
\>\>\>                     search\_setup(Vars, IdMap, Level)\\
\>\>                 )\\
\>             ).\\
\\
search([], \_, \_).\\
search([Var-Why | VarWhys], IdMap, Level) :-\\
\>	NextLevel is Level + 1,\\
\>	catch(\\
\>\>			(\\
\>\>\>				bind(Var-Why, Level),\\
\>\>\>				search(VarWhys, IdMap, NextLevel) \\
\>\>			),\\
\>\>			ball(Level, Clause),\\
\>\>			(\\
\>\>\>                                update\_conflict(NextLevel, Clause, Level, IdMap),\\
\>\>\>				search([Var-Why | VarWhys], IdMap, NextLevel)\\
\>\>			)\\
\>	     ).\\
\\
bind(Var-Why, Level) :-\\
\>        (\\
\>\>            var(Var)\\
\>        ->\\
\>\>            Var-Why = true-imp(Level-0, \_, true, [])\\
\>	;\\
\>\>            true \\
\>        ).\\
\\
conflict(Whys) :- \\
\>	analyse\_conflict(Whys, BackjumpLevel, Clause), \\
\>        throw(ball(BackjumpLevel, Clause)). \\
\end{tabbing}
\caption{Backjumping search for SAT}
\label{fig:backjump}
\end{figure}

\begin{figure}[t!]
\tt
\begin{tabbing}
4ex\=4ex\=4ex\=4ex\=4ex\=4ex\=4ex\kill
update\_conflict(NextLevel, Clause, Level, IdMap) :-\\
\>	get\_learnt(Learnt),\\
\>	Learnt1 = [NextLevel-Clause | Learnt],\\
\>	add\_learnt\_clauses(Learnt1, Level, IdMap),\\
\>	put\_learnt(Learnt1).\\
\\
add\_learnt\_clauses([], \_, \_).\\
add\_learnt\_clauses([Level-Clause | Clauses], DecisionLevel, IdMap) :-\\
\>	(\\
\>\>		Level < DecisionLevel\\
\>	->\\
\>\>		true\\
\>	;\\
\>\>		unground(Clause, IdMap, NewClause),\\
\>\>		watch\_clause(NewClause)\\
\>	),\\
\>	add\_learnt\_clauses(Clauses, DecisionLevel, IdMap).\\
\\
unground([], \_, []).\\
unground([Pol-Id | Labels], IdMap, [Pol-VarImp | Literals]) :- \\
\>	get\_assoc(Id, IdMap, VarImp),\\
\>	unground(Labels, IdMap, Literals).\\
\\
get\_learnt(Learnt) :- bb\_get(learnt, Learnt). \\
put\_learnt(Learnt) :- bb\_put(learnt, Learnt). \\
\end{tabbing}
\caption{Conflict analysis}
\label{fig:conflict}
\end{figure}

\paragraph{Search with backjumping for SAT}
Figure~\ref{fig:backjump} gives code which realises
backjumping search for SAT.  The main search predicate is
\verb|search(VarWhys, IdMap, Level)| where, like before,
\verb|VarWhys| is a list of pairs, but here each pair is a variable conjoined with
an implication graph which explains its binding. Search is controlled
by overlaying backjumping with learning, in which
the reason for a conflict is summarised by a conjunction of propositional
literals whose negation gives a clause that is implied by the SAT instance. The clause (often referred to as a blocking clause)
is then added to the problem to steer search away from the conflict.
The map \verb|IdMap| associates a ground identifier with the variable it represents, 
needed to reconstruct a (non-ground) clause from a conflict.
\verb|Level| is the decision level which a variable adopts if
it is assigned by search rather than propagation. The first decision level is taken to be 1 
(though, as discussed later, decision level 0 can also occur).

Consider first the second clause of \verb|search| which is responsible
for backjumping, learning and labelling.
Each \verb|catch| meta-call is concerned with
assigning one variable if it is unassigned.
The predicate \verb|bind| realises
labelling, and instantiates \verb|Var| to a truth value,
whilst recording the value and level in a (leaf) node of an implication graph.
Note that \verb|bind| is never backtracked into, and assignment to \verb|false| is
not explicitly made.
Since this code is realising CDCL, failure of a binding will lead to a 
blocking clause 
being learnt and added to the problem description which will guide search away from the assignment that has failed, directing search to the \verb|false| 
branch.  
If a conflict is discovered during search, the \verb|set_watch| predicate in Fig.~\ref{fig:difSetup}
will call \verb|conflict(Whys)| in Fig.~\ref{fig:backjump} and a
ball is created with \verb|BackjumpLevel|
instantiated to the decision level of the backjump. The throw will thus only match
the unique \verb|Catcher| term \verb|ball(Level, Clause)| for which \verb|Level = BackjumpLevel|.
The ground \verb|BackjumpLevel| and \verb|Clause| terms are formed by the predicate 
\verb|analyse_conflict(Whys, BackjumpLevel, Clause)|
using a conflicting list of implication graphs \verb|Whys| discovered when watching a clause (see Section~\ref{sec:learn} and Fig~\ref{fig:reinstate}).
On recovery, the bindings made by \verb|search| and
propagation are unwound back to the backjump level as required (though
the bindings on the ball are retained).
The \verb|RecoveryGoal| then uses \verb|update_conflict| in Fig.~\ref{fig:conflict} to add \verb|Clause| to a database of
learnt clauses (the learning algorithm explained in Section~\ref{sec:learn})
which grows as search proceeds.  The ground \verb|Clause| is translated into
its non-ground representation at the stage of the \verb|RecoveryGoal| where the
decision level reverts to that of \verb|BackjumpLevel|:
the variables of the clause can have bindings at later decision levels
which do not hold at the \verb|BackjumpLevel|.
Other elements of the database of learnt clause may also need to be restored 
in a similar way.

The predicates \verb|get_learnt(Learnt)| and \verb|put_learnt(Learnt)| are merely
wrappers to builtins that read and
write a list of (learnt) clauses to a non-backtrackable database: they
can be realised with \verb|bb_get|/\verb|bb_put| as illustrated in Fig.~\ref{fig:conflict}, or 
\verb|nb_getval|/\verb|nb_setval| \cite{wielemaker11swi}, or using the assert/retract family of built-ins.

When considering graph colouring in Section~\ref{sec:colour}, colours are always explicitly
assigned in a search phase, and a conflict can only occur 
during search, after a colour is assigned.  
However, SAT is different: a variable might
be assigned a value before search at the initial set up.
The decision level of such an assignment is taken to be 0.
Conflicts which arise at this level are handled by the singleton case in \verb|analyse_conflict|,
which immediately fails without reaching a \verb|throw|, since if the
problem specification is unsatisfiable then no further search is necessary.  
Furthermore, with assignment at 
decision level 0 possible, conflict analysis might determine that level 0
is the appropriate decision level to jump to.  This is prior to the call to \verb|search|, 
hence the addition of \verb|search_setup| to handle these backjumps; this predicate
mirrors \verb|search|, but without the call to \verb|bind|.  In addition, it is 
possible that the backjump caught by \verb|ball(0, Clause)| describes a unit \verb|Clause| 
and this itself leads to assignments at decision level 0.


\subsection{Conflict Analysis for SAT}\label{sec:learn}

When a conflict is encountered, the implication graph leading to it is examined
to learn a clause, which is added to the formula to steer search away from the
conflict.  To illustrate, consider the formula
$f = (\neg x_1 \vee x_8 \vee \neg x_2) \wedge
(\neg x_1 \vee \neg x_3) \wedge
(x_2 \vee x_3 \vee x_4 ) \wedge
(\neg x_4 \vee \neg x_5) \wedge
(x_5 \vee x_6) \wedge
(x_7 \vee \neg x_4 \vee \neg x_6)$ and the partial assignment
$\theta = \{ x_7 \mapsto \mathit{false}, x_8 \mapsto \mathit{false} \}$
where the variables  $x_7$ and $x_8$ are assigned at decision levels 1 and 2
respectively by search.  Observe that no further bindings are
inferred by unit propagation.  However, if search subsequently adds
$x_1 \mapsto \mathit{true}$ at decision level 3 then a series of unit
propagations ensue that ultimately lead to a conflict, owing to
unsatisfiability of the clause $(x_7 \vee \neg x_4 \vee \neg x_6)$.

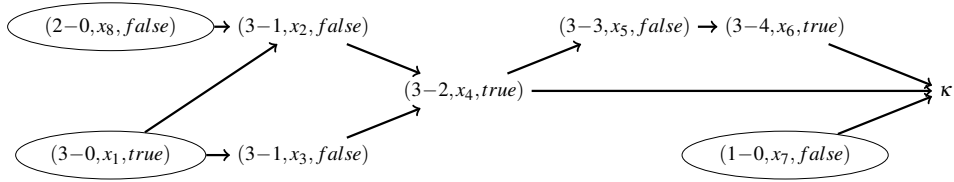
\begin{figure}[t!]
\centering

\newcommand{\spacing}{2.5}
\resizebox{\textwidth}{!}
{\begin{tikzpicture}
\usetikzlibrary{arrows}
\usetikzlibrary{shapes}
    
\node[draw, ellipse] at (-0.5, 2) (x8){\small{$(2\!-\!0, x_8, false)$}};
\node[draw, ellipse] at (-0.5, 0) (x1){\small{$(3\!-\!0, x_1, true)$}};   
\node at (1 * \spacing, 2) (x2){\small{$(3\!-\!1, x_2, false)$}};
\node at (1 * \spacing, 0) (x3){\small{$(3\!-\!1, x_3, false)$}};
\node at (2 * \spacing, 1) (x4){\small{$(3\!-\!2, x_4, true)$}};
\node at (3 * \spacing, 2) (x5){\small{$(3\!-\!3, x_5, false)$}};
\node at (4 * \spacing, 2) (x6){\small{$(3\!-\!4, x_6, true)$}};
\node[draw, ellipse] at (4 * \spacing, 0) (x7){\small{$(1\!-\!0, x_7, false)$}};
\node at (5 * \spacing, 1) (kappa){\small{$\kappa$}};

\foreach \i/\j in {
      x8/x2,
      x1/x3,
      x1/x2,
      x2/x4,
      x3/x4,
      x4/kappa,
      x4/x5,
      x5/x6,
      x7/kappa,
      x6/kappa}
\draw [arrows=->, ,line width=1pt] (\i) --  (\j);

\end{tikzpicture}}
\vspace*{6pt}
\caption{Implication graph from $f$ and 
the partial assignment
$\theta = \{ x_7 \mapsto \mathit{false}, x_8 \mapsto \mathit{false}, x_1 \mapsto \mathit{true} \}$}
\label{fig:imp}
\end{figure}

\paragraph{Unique Implication Points} 
Figure~\ref{fig:imp} illustrates an implication graph rooted at a special
conflict node,
$\kappa$, that details how the conflict follows from binding $x_4$, $x_6$ and $x_7$.
Each triple in the figure gives the level and sublevel,
the variable assigned, and its truth value.
The implication graph can be inspected to learn a clause. The leaves of the
implication graph, which are circled in Fig.~\ref{fig:imp}, are the three
bindings $x_7 \mapsto \mathit{false}$, $x_8 \mapsto \mathit{false}$ and $x_1
\mapsto \mathit{true}$, that together with clauses of $f$, prohibit the
satisfiability of $(x_7 \vee \neg x_4 \vee \neg x_6)$.  This combination of
bindings can be avoided by adding $(x_7 \vee x_8 \vee \neg x_1)$ to $f$ which
is therefore a candidate learnt clause.  However, another choice is possible,
notably one with fewer literals.  Observe that any path that starts with the binding
of $x_1$ at decision level 3, the current decision level, and
ends at $\kappa$ passes through the intermediate node
where $x_4$ is assigned.  The single binding $x_4 \mapsto \mathit{true}$
therefore summarises the net effect of the two bindings $x_1 \mapsto
\mathit{true}$ and $x_8 \mapsto \mathit{false}$.  Thus, an alternative learnt
clause is $(\neg x_4 \vee x_7)$.
In general, any node between the current (most recent) decision variable and the
conflict $\kappa$ that strictly dominates \cite{simplefast06cooper} $\kappa$ can
be used to construct a
learnt clause.  Such nodes are termed Unique Implication Points (UIPs). The
UIP nearest $\kappa$ is the first UIP (the node
where $x_4$ is assigned). The last UIP is furthest from $\kappa$ and is
where the current decision variable is bound (the node
where $x_1$ is assigned).

\begin{figure}[t!]
\tt
\begin{tabbing}
4ex\=4ex\=4ex\=4ex\=4ex\=4ex\=4ex\kill
analyse\_conflict(Whys, BackjumpLevel, Clause) :- \\
\>	construct\_clause(Whys, Lits, Levels), \\
\>	remove\_dups(Lits, Clause), \\
\>	sort(Levels, SortedLevels), \\
\>	( \\
\>\>		SortedLevels = [BackjumpLevel] \\
\>	-> \\
\>\>		BackjumpLevel \notunify 0 \\
\>	; \\
\>\>		reverse(SortedLevels, [\_, BackjumpLevel | \_]) \\
\>	). \\
\\
construct\_clause([], [], []). \\
construct\_clause([imp(Level, Id, Pol, SubWhys) \guillotine\ Whys], Lits, Levels) :- \\
\> ( \\
\> \> SubWhys == [] \\
\> -> \\
\> \> negate(Pol, NegPol), \\
\> \> Lits = [NegPol-Id \guillotine\ MoreLits], \\
\> \> Levels = [Level \guillotine\ MoreLevels], \\
\> \> construct\_clause(Whys, MoreLits, MoreLevels) \\
\> ; \\
\> \> append(SubWhys, Whys, AllWhys), \\
\> \> construct\_clause(AllWhys, Lits, Levels) \\
\> ). \\
\\
negate(true, false). \\
negate(false, true). \\
\end{tabbing}
\caption{Inferring and instantiating learnt clauses}
\label{fig:reinstate}
\end{figure}

Figure~\ref{fig:reinstate} presents code for creating a learnt clause
based on the last UIP. The predicate \verb|construct_clause| performs a
depth-first traversal of the implication graph, starting from
$\kappa$, and identifies decision variables by their empty implication graphs
(\verb|SubWhys|).  The constructed clause is made up of literals where
variables are identified by the \verb|Id| ground term, and
polarities are the negation of those that caused the conflict.

The \verb|construct_clause| predicate also gathers the decision levels at which
literals in the new clause were assigned. These levels are used to find
the backjump level, which is chosen to be such that the learnt
clause becomes unit, directing search away from the conflict.
For the last UIP learnt clause $x_7 \vee x_8 \vee \neg x_1$
of Fig.~\ref{fig:imp},  \verb|construct_clause| will
derive the clause
\verb|[false-x1, true-x8, true-x7]| with decision levels \verb|[3, 2, 1]|.  
Search
should resume by backjumping to level 2.  
Therefore the call to \verb|conflict(Whys)| leads to
\verb|throw(ball(2, [false-x1, true-x8, true-x7]))|.
More generally, the backjump
level is the largest level strictly less than the maximum (which is actually
the current decision level); search would not immediately benefit from the new
clause if resumed at an earlier decision level. Continuing at the backjump level
ensures that the learnt clause becomes unit almost immediately.

The act of backjumping removes bindings, in particular those induced by recently
added clauses. However, these clauses are retained in the non-backtrackable database 
using a ground representation. Fig.~\ref{fig:conflict} lists the code for reinstating 
learnt clauses after backjumping using the predicate \verb|add_learnt_clauses|.  Learnt clauses are
saved together with the level at which they were learned.
Upon
backjumping,
the calls to \verb|watch_clause| for clauses learnt at the backjump
decision level or later are lost, and these calls need to be made again.
This necessitates rebuilding clauses after backjumping from the ground representation
of the database. The \verb|unground|
predicate achieves this using association list \verb|IdMap| to map a ground identifier
to its propositional variable. This list is built
during problem setup and is passed through \verb|search| as shown in
Fig.~\ref{fig:backjump}.

Reinstatement of learned clauses can be performed selectively to realise
$k$-learning. In $k$-learning \cite{marquessilva96grasp}, 
learned clauses are only added to the constraint store permanently
if they have less than $k$ variables. It decreases the cost of learning by lessening pressure on memory and reducing the number of updates to the store of clauses.  Clauses with fewer variables
have most influence, but learning will have little impact if $k$ is set too low.
The presented approach can be also modified to allow first UIP clause learning. This
typically produces smaller clauses more tightly focused on the cause of the
conflict~\cite{marquessilva09cdcl}.  First UIP can be realised in Prolog
by running a frontier over the implication graph, starting at $\kappa$, repeatedly
expanding the node with the highest level and sublevel. The first UIP is found when the frontier reduces to contain a
single node at the highest level (possibly augmented with nodes of lower level).
First UIP can thus be found in a single pass over the conflict graph
without introducing additional data structures.  

Backjumping is designed to accelerate search which begs the question
of whether backjumping, when implemented with \verb|catch| and \verb|throw|, can
ever improve on the default search mode of Prolog.
Table~\ref{tbl:results} presents timings for a biased sample of classic SAT benchmarks;
biased because they were chosen to be non-trivial in that
the execution time exceeds 2~seconds for the vanilla solver. 
Times are in
milliseconds, and benchmarking was carried out with SICStus 4.5.1 on a
2.5GHz Macbook Pro with 16GB RAM. 
The first 2 benchmarks
are random 3-SAT instances with controlled backbone size,
the next 4 originate from flat graph colouring problems, and
the final 4 include 2
unsatisfiable and 2 satisfiable random 3-SAT instances.
The table gives data for the vanilla solver with no learning, and for
the backjumping solver given in this paper, augmented with first UIP conflict
driven clause learning and using
$k$-learning with $k=8$.  For both solvers, the execution time in milliseconds
is given (time), alongside the number of variable assignments made (assign).  
In addition, for the 
backjumping solver, the number of throws made by the solver (throws), and the number of assignments jumped over (jumps) are given.
The balls thrown in backjumping are the size of the analysed conflict which is
typically less than 20 literals for these benchmarks, though this is problem dependent.
The timings suggest that \verb|catch| and \verb|throw| are not only a useful
code structuring device, but can enable significant improvement in performance
when 
used to realise backjumping. 
As is conventional in SAT solving, the timings are for finding a
single solution.  To enumerate all solutions, the template from 
graph colouring using the \verb|succeed| predicate can be adapted
to add a blocking clause for the solution and reactivating search
reusing the learnt clauses.
Code for the work presented in this paper
is available at \url{https://www.cs.kent.ac.uk/~amk/backjump.zip},
including some variations not discussed here
such as pruning the set of learnt clauses by removing those which are entailed
by a newly learnt clause.


\begin{table}[t!]
\centering
\caption{Performance of learning versus no learning}
\begin{tabular}{@{}l|@{}r @{}r | @{}r @{}r @{}r @{}r }
& \multicolumn{2}{c|}{no learning} & \multicolumn{4}{c}{$8$-learning} \\
\cline{1-7} 
benchmark & time & assign & time & assign & throws & jumps \\
\cline{1-7}
CBS\_k3\_n100\_m435\_b90\_127 & 2987 & 179427 & 2206 & 56329 & 24650 & 31667 \\
CBS\_k3\_n100\_m435\_b90\_139 & 3204 & 193247 & 874 & 26827 & 10163 & 16647 \\
flat175-17 & 22791 & 748377 & 10225 & 107066 & 43546 & 63477 \\
flat175-28 & 15754 & 471521 & 15132 & 154985 & 70895 & 84059 \\
flat200-20 & 16345 & 519868 & 3252 & 39674 & 16107 & 23524 \\
flat200-39 & $>$30000 &  & 13326 & 125053 & 54516 & 70502 \\
uf100-0126 & 4097 & 248581 & 1984 & 53320 & 23009 & 30293 \\
uf100-015 & 3499 & 210330 & 857 & 24676 & 10101 & 14559 \\
uuf100-0119 & 9651 & 634568 & 3644 & 95553 & 40073 & 55497 \\
uuf100-0120 & 5731 & 350866 & 2578 & 67868 & 28378 & 39498 \\
\end{tabular}
\label{tbl:results}
\end{table}


\section{Concluding Discussion} \label{sec:discuss}

The idea of using catch and throw to realise backjumping dates
back at least to a discussion \cite{baljeu05backjump} on \verb|comp.lang.prolog|
but the authors are not aware of any 
programming examples that serve to illustrate the technique.
The message of this discussion is exemplified with two case
studies which demonstrate that \verb|catch| and \verb|throw| 
are more versatile than one would expect,
providing exactly what is required for programming backjumping. In fact,
random solver restarts \cite{howe12pearl} can also be accommodated by adding an outer \verb|catch|
meta-call which intercepts restart exceptions thrown
when the number of backjumps exceeds a threshold.

Intelligent backtracking \cite{bruynooghe04enchancing} can be combined with learning
by restarting the search from scratch and then fast-forwarding to a
particular decision point  \cite{howe12pearl}. 
However, \verb|catch| and \verb|throw| provide a more general
solution, not requiring search to be completely restarted,
and there seems no reason why this strategy cannot extend to SMT solving \cite{robbins15theory} by
another instantiation of the template.
It has been recently shown \cite{drabent18logic} that the vanilla SAT solver of Fig.~\ref{fig:sat-solver} can be
understood as a logic program with added control, however 
reasoning about the correctness of learnt clauses is more challenging still,
a research problem that is not exclusive to logic programming.

Realising backjumping with an ISO feature is undoubtedly attractive, 
and since ISO encourages the use of catch and throw, one
might hope that increasingly efficient implementations will emerge over time.
Even though SWI-Prolog does not comply with the ISO specification
(since it does not copy the \verb|Ball| before unifying it with \verb|Catcher|)
the two case studies are fully portable because 
the \verb|Goal| and \verb|Catcher| goals do not share variables.


\paragraph{Acknowledgments}
The authors thank the editor and anonymous referees for their thought-provoking comments.  This is a better paper for these suggestions.
This work was supported, in part, by EPSRC grants EP/K031929/1 and EP/N020243/1.


\bibliographystyle{acmtrans}
\bibliography{pearl}

\label{lastpage}
\end{document}